\documentclass[11pt,twoside]{article}
\usepackage{asp2010}

\resetcounters

\begin{document}

\title{Map-making for large-format detector arrays on CCAT}
\author{Gaelen~Marsden$^1$, Tim~Jenness$^2$, and Douglas~Scott$^1$
\affil{$^1$Department of Physics and Astronomy, University of British
  Columbia, 6224~Agricultural~Road, Vancouver, BC V6T~1Z1, Canada}
\affil{$^2$Department of Astronomy, Cornell University, Ithaca NY,
  14853, USA}}

\begin{abstract}
CCAT is a large submillimetre telescope to be built near the ALMA
site in northern Chile. A large-format KID camera, with up to 48,000
detectors at a single waveband sampled at ${\sim}1$\,kHz, will have a
data rate ${\sim}50$ times larger than SCUBA-2, the largest existing
submillimetre camera. Creating a map from this volume of data will be
a challenge, both in terms of memory and processing time required. We
investigate how to extend SMURF, the iterative map-maker used for
reducing SCUBA-2 observations, to a distributed-node parallel system,
and estimate how the processing time scales with the number of nodes
in the system.
\end{abstract}

\section{Introduction}

CCAT \citep[see e.g.][]{2014SPIE9152-109} will be a large telescope
optimized for observations in the submillimetre and will be located
near the ALMA site in northern Chile. One of the proposed instruments
is SWCam \citep{Stacey2014}, a large-format kinetic inductance
detector (KID) array with subarrays spanning 850--350\,\micron, and
possibly extending to 200\,\micron. With a field of view of 16~arcmin
and a sampling rate of ${\sim}1\,$kHz, SWCam will be able to quickly
survey large fields (tens of square degrees) at high resolution
(3.5\arcsec\ at 350\,\micron) and sensitivity; it will be able to
reach noise levels five times below the confusion limit at each
wavelength over 35\,deg$^2$ in the first year of observations.

Current submillimetre photometric instruments, such as SHARC-II
\citep{2003SPIE.4855...73D}, LABOCA \citep{2009A&A...497..945S},
ArT\'{e}MiS \citep{reveret2014} and SCUBA-2
\citep{2013MNRAS.430.2513H}, have hundreds to thousands of detectors
sampled at tens to hundreds of Hertz. The SWCam data volume of
5~TB/night will thus be considerably larger than current data sets,
and we must ensure that map-making software will be able to reduce the
data in a reasonable amount of time. \cite{2014ASPC..485..399M}
discuss the problem of scaling current map-makers to SWCam, and
suggest distributed-memory parallelization as a solution to handling
the large data sets. Here, we further develop the distributed-memory
parallel model and present results of timing tests.

\section{Distributed Memory Parallel Map-making}

\begin{figure}[ht]
\centering
\includegraphics[width=\textwidth]{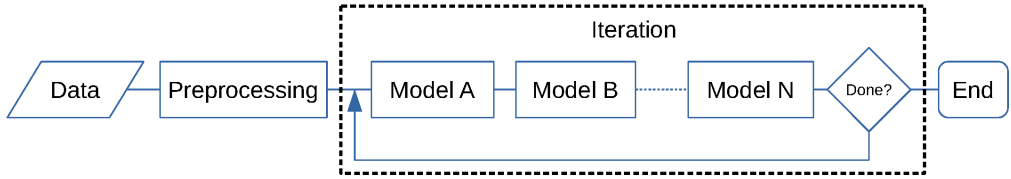}
\caption{Flow chart illustrating iterative map-making using $N$
  models. The models represent sources of signal and noise, such as
  common-mode signal,
  low frequency $1/f$ noise, and astronomical signal.
\label{fig:flowchart}}
\end{figure}

Following \cite{2014ASPC..485..399M}, we focus on SMURF
\citep[][ascl:1310.007]{2013MNRAS.430.2545C}, the iterative map-maker used for
reduction of SCUBA-2 data. A simplified flow chart describing the
algorithm is shown in Figure~\ref{fig:flowchart}.
Very briefly, the data are read from disk, preprocessing steps
(e.g.\ spike detection) are performed, then a number of noise/signal
models are fit and removed from the data.\footnote{The number, type
  and sequence of models fit is a configuration option.} At the end of
the sequence, the r.m.s.\ of the time stream residuals is compared to the
previous iteration; if the change is larger than the convergence
tolerance, the sequence of models is re-fit to the new residuals. The
iteration loop continues until the convergence condition is met, up to
a maximum number of iterations.

\begin{figure}[ht]
\centering
\includegraphics[width=\textwidth]{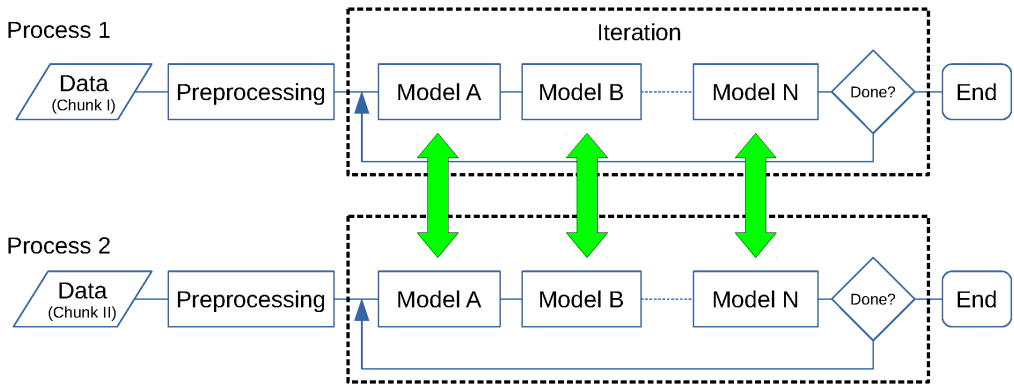}
\caption{Distributed-memory parallel version of the iterative
  map-maker. Each process is responsible for a distinct chunk of
  data. Communication between processors is necessary for most (but
  not all) models. The two-processor version is shown---this is easily
  extended to $M$ processors. Hierarchical communication between all
  processors is required.
\label{fig:flowchart2}}
\end{figure}

\cite{2014ASPC..485..399M} suggest that, even accounting for advances
in computer hardware in the next five years, a single machine will not
be able reduce more than 15 minutes of data at a time, and will not be
able to keep up with data collection. Adapting the iterative algorithm
to run on a distributed-memory parallel computer would allow for the
reduction of data sets longer than 15 minutes (necessary for making
high-fidelity maps) and for the reduction run
time. Figure~\ref{fig:flowchart2} shows how this can be accomplished.
For $M$ processors, we divide the data set into $M$ chunks, with each
processor handling only its chunk of data. Communication between the
processes is required when calculating models that depend on the full
data set. This communication is represented by the large green arrows
in Figure~\ref{fig:flowchart2}. Other than these communication steps,
the processors run independently.

For the most part, only (relatively) small data arrays need to be
shared. As an example, the common mode model (a time-varying
spatially-independent signal seen by all detectors), each process
accumulates the sum of signal over detectors in its data chunk, then
the accumulated sums are combined over all processes using collective
communication. For the astronomical sky model (in other words, the sky
map), each process accumulates signal into its own copy of the sky map
and the maps are shared in the same manner as for the common mode
model. For models such as the high-pass filter (which reduces
low-frequency detector noise and atmospheric noise), no communication
is necessary, since the result for each detector is independent of all
others.

\section{Prototype and Timing Tests}

We have written a prototype map-maker\footnote{Available at:
  \url{https://github.com/CCATObservatory/mpi-mapmaker-test}.} to test
the parallel iterative algorithm described above. The code is
written in C and uses \texttt{Open MPI}.\footnote{An open source implementation
  of the Message Passing Interface. See:
  \url{http://www.open-mpi.org/}.} The prototype implements the three
models discussed: common mode, high-pass filter and astronomical
signal.

\begin{figure}[ht]
\centering
\includegraphics[width=0.8\textwidth]{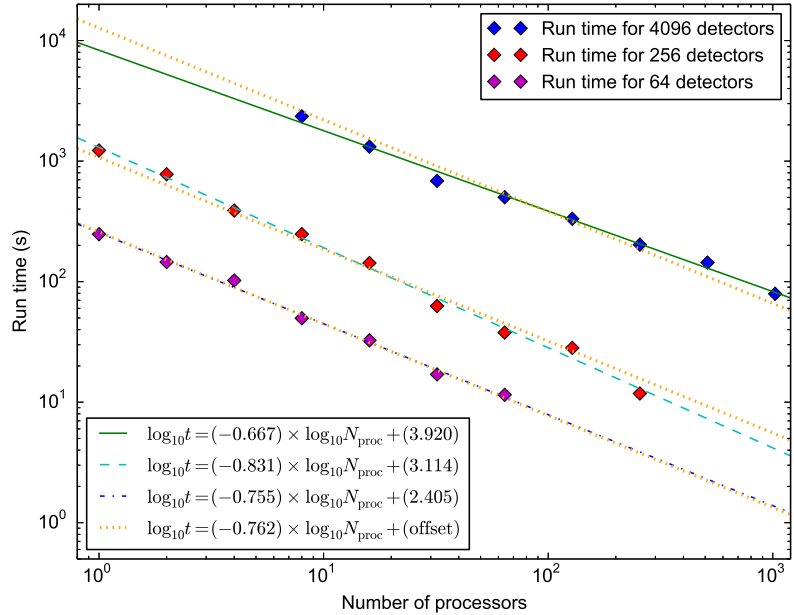}
\caption{Timing tests for the prototype parallel iterative
  map-maker. Run time vs.\ number of processors for three simulated
  data sets is shown. Each data set is 30 minutes of data sampled at
  1500\,Hz for the indicated number of detectors. The run times show
  power-law scaling. A power-law fit to each data set is shown, as is
  a fit to all three data sets with a common power law index (orange
  dotted curves).
\label{fig:times}}
\end{figure}

The prototype has been run on a range of data set sizes and a varying
number of processors in order to determine how the run time scales
with the number of processors. The results of these tests are shown in
Figure~\ref{fig:times}.
The scaling of run times with number of processors for each data set
size is well-described by a power law, and all three are reasonably
well-fit by a single power-law index (shown in the figure as
orange dotted lines), $\alpha=-0.762$. This corresponds to a reduction
of run time by a factor of 1.7 for each factor of two in the number of
processors, or a factor of ${\sim}20$ for 50 processors. This is
sufficient to reduce the run time of SWCam data to better than rate at
which the data are collected.

\section{Conclusions}

We have demonstrated that the iterative map-maker used by SCUBA-2 can
be adapted for use on a distributed-memory parallel system. A
relatively modest cluster of 50 processors will be able to reduce
SWCam data faster than they are collected, and reduces the memory
required per processor to ${\sim}200$\,GB per hour of data. We are now
adapting SMURF so that it can be run on such a system.

\acknowledgements Thanks to Jack Sayers for providing the data
simulation code that was used as the input to the timing tests. The
SWCam prototype map-maker makes use of facilities provided by WestGrid
and Compute Canada Calcul
Canada.\footnote{\url{http://www.computecanada.ca}}

\bibliography{P2-6}

\end{document}